\documentclass[ALICE,manyauthors]{cernphprep}
\usepackage[comma,square,numbers,sort&compress]{natbib}
\usepackage{hyperref}
\usepackage{lineno}
\usepackage{siunitx}
\DeclareSIUnit\clight{\text{\ensuremath{c}}}
\usepackage{xspace}
\usepackage{bm}
\usepackage[T1]{fontenc}
\begin{document}
%

\newcommand{\pp}           {pp\xspace}
\newcommand{\ppbar}        {\mbox{$\mathrm {p\overline{p}}$}\xspace}
\newcommand{\XeXe}         {\mbox{Xe--Xe}\xspace}
\newcommand{\PbPb}         {\mbox{Pb--Pb}\xspace}
\newcommand{\pPb}          {\mbox{p--Pb}\xspace}
\newcommand{\AuAu}         {\mbox{Au--Au}\xspace}
\newcommand{\dAu}          {\mbox{d--Au}\xspace}

\newcommand{\roots}        {\ensuremath{\sqrt{s}}\xspace}
\newcommand{\snn}          {\ensuremath{\sqrt{s_{\mathrm{NN}}}}\xspace}
\newcommand{\pt}           {\ensuremath{p_{\rm T}}\xspace}
\newcommand{\meanpt}       {$\langle p_{\mathrm{T}}\rangle$\xspace}
\newcommand{\ycms}         {\ensuremath{y_{\rm CMS}}\xspace}
\newcommand{\ylab}         {\ensuremath{y_{\rm lab}}\xspace}
\newcommand{\etarange}[1]  {\mbox{$\left | \eta \right |~<~#1$}}
\newcommand{\yrange}[1]    {\mbox{$\left | y \right |~<~#1$}}
\newcommand{\dndy}         {\ensuremath{\mathrm{d}N_\mathrm{ch}/\mathrm{d}y}\xspace}
\newcommand{\dndeta}       {\ensuremath{\mathrm{d}N_\mathrm{ch}/\mathrm{d}\eta}\xspace}
\newcommand{\avdndeta}     {\ensuremath{\langle\dndeta\rangle}\xspace}
\newcommand{\dNdy}         {\ensuremath{\mathrm{d}N_\mathrm{ch}/\mathrm{d}y}\xspace}
\newcommand{\Npart}        {\ensuremath{N_\mathrm{part}}\xspace}
\newcommand{\Ncoll}        {\ensuremath{N_\mathrm{coll}}\xspace}
\newcommand{\dEdx}         {\ensuremath{\textrm{d}E/\textrm{d}x}\xspace}
\newcommand{\RpPb}         {\ensuremath{R_{\rm pPb}}\xspace}
\newcommand{\kstar}        {\ensuremath{k^*}\xspace}
\newcommand{\mt}           {\ensuremath{m_{\rm{T}}}\xspace}

\newcommand{\nineH}        {$\sqrt{s}~=~0.9$~Te\kern-.1emV\xspace}
\newcommand{\seven}        {$\sqrt{s}~=~7$~Te\kern-.1emV\xspace}
\newcommand{\thirteen}     {$\sqrt{s}~=~13$~Te\kern-.1emV\xspace}
\newcommand{\forteen}     {$\sqrt{s}~=~14$~Te\kern-.1emV\xspace}
\newcommand{\twoH}         {$\sqrt{s}~=~0.2$~Te\kern-.1emV\xspace}
\newcommand{\twosevensix}  {$\sqrt{s}~=~2.76$~Te\kern-.1emV\xspace}
\newcommand{\five}         {$\sqrt{s}~=~5.02$~Te\kern-.1emV\xspace}
\newcommand{\fivefive}     {$\sqrt{s}~=~5.5$~Te\kern-.1emV\xspace}
\newcommand{\twosevensixnn}{$\sqrt{s_{\mathrm{NN}}}~=~2.76$~Te\kern-.1emV\xspace}
\newcommand{\fivenn}       {$\sqrt{s_{\mathrm{NN}}}~=~5.02$~Te\kern-.1emV\xspace}
\newcommand{\LT}           {L{\'e}vy-Tsallis\xspace}
\newcommand{\GeVc}         {GeV/$c$\xspace}
\newcommand{\MeVc}         {MeV/$c$\xspace}
\newcommand{\tev}          {TeV\xspace}
\newcommand{\gev}          {GeV\xspace}
\newcommand{\mev}          {MeV\xspace}
\newcommand{\GeVmass}      {GeV/$c^2$\xspace}
\newcommand{\MeVmass}      {MeV/$c^2$\xspace}
\newcommand{\lumi}         {\ensuremath{\mathcal{L}}\xspace}

\newcommand{\ITS}          {\rm{ITS}\xspace}
\newcommand{\TOF}          {\rm{TOF}\xspace}
\newcommand{\ZDC}          {\rm{ZDC}\xspace}
\newcommand{\ZDCs}         {\rm{ZDCs}\xspace}
\newcommand{\ZNA}          {\rm{ZNA}\xspace}
\newcommand{\ZNC}          {\rm{ZNC}\xspace}
\newcommand{\SPD}          {\rm{SPD}\xspace}
\newcommand{\SDD}          {\rm{SDD}\xspace}
\newcommand{\SSD}          {\rm{SSD}\xspace}
\newcommand{\TPC}          {\rm{TPC}\xspace}
\newcommand{\TRD}          {\rm{TRD}\xspace}
\newcommand{\TZERO}        {\rm{T0}\xspace}
\newcommand{\VZERO}        {\rm{V0}\xspace}
\newcommand{\VZEROA}       {\rm{V0A}\xspace}
\newcommand{\VZEROC}       {\rm{V0C}\xspace}
\newcommand{\Vdecay} 	   {\ensuremath{V^{0}}\xspace}

\newcommand{\phot}         {\ensuremath{\gamma}\xspace} 
\newcommand{\ee}           {\ensuremath{\mathrm{e}^{+}\mathrm{e}^{-}}\xspace} 
\newcommand{\pip}          {\ensuremath{\uppi^{+}}\xspace}
\newcommand{\pim}          {\ensuremath{\uppi^{-}}\xspace}
\newcommand{\kap}          {\ensuremath{\rm{K}^{+}}\xspace}
\newcommand{\kam}          {\ensuremath{\rm{K}^{-}}\xspace}
\newcommand{\pbar}         {\ensuremath{\rm\overline{p}}\xspace}
\newcommand{\kzero}        {\ensuremath{{\rm K}^{0}_{\rm{S}}}\xspace}
\newcommand{\lmb}          {\ensuremath{\upLambda}\xspace}
\newcommand{\sigm}         {\ensuremath{\upSigma}\xspace}
\newcommand{\almb}         {\ensuremath{\overline{\upLambda}}\xspace}
\newcommand{\Om}           {\ensuremath{\upOmega^-}\xspace}
\newcommand{\Mo}           {\ensuremath{\overline{\upOmega}^+}\xspace}
\newcommand{\X}            {\ensuremath{\upXi^-}\xspace}
\newcommand{\Ix}           {\ensuremath{\overline{\upXi}^+}\xspace}
\newcommand{\Xis}          {\ensuremath{\upXi^{\pm}}\xspace}
\newcommand{\Oms}          {\ensuremath{\upOmega^{\pm}}\xspace}
\newcommand{\degr}         {\ensuremath{^{\rm o}}\xspace}

\newcommand{\siZ}          {\ensuremath{\upSigma^{0}}\xspace}
\newcommand{\siP}          {\ensuremath{\upSigma^{+}}\xspace}
\newcommand{\siM}          {\ensuremath{\upSigma^{-}}\xspace}
\newcommand{\asiZ}         {\ensuremath{\overline{\upSigma^{0}}}\xspace}
\newcommand{\psiZCF}       {\rm{p}\mbox{--}\siZ}
\newcommand{\SNCF}         {\rm{N}\mbox{--}\ensuremath{\upSigma}\xspace}
\newcommand{\LNCF}         {\rm{N}\mbox{--}\ensuremath{\upLambda}\xspace}
\newcommand{\apasiZCF}     {\ensuremath{\overline{\mathrm{p}}}\mbox{--}\asiZ}
\newcommand{\psiZCombCF}   {\rm{p}\mbox{--}\siZ \ensuremath{\oplus} \apasiZCF}
\newcommand{\SB}           {\ensuremath{(\upLambda\gamma)}\xspace}
\newcommand{\pSBCF}        {\rm{p}\mbox{--}\SB}

\newcommand{\pLCF}         {\rm{p}\mbox{--}\lmb}
\newcommand{\nSplusCF}     {\rm{n}\mbox{--}\ensuremath{\upSigma^{+}}}

\newcommand{\Ledn}         {Lednick\'y--Lyuboshits approach\xspace}
\newcommand{\chiEFT}       {\ensuremath{\chi}\rm{EFT}\xspace}
\newcommand{\fss}       {\rm{fss2}\xspace}
\newcommand{\ESC}          {\rm{ESC16}\xspace}
\newcommand{\NSC}          {\rm{NSC97f}\xspace}

\newcommand{\pP}           {\ensuremath{\mbox{p--p}}\xspace}
\newcommand{\ApAP}         {\ensuremath{\mbox{\pbar--\pbar}}\xspace}
\newcommand{\pPComb}       {\ensuremath{\mbox{p--p} \oplus \mbox{\pbar--\pbar}}\xspace}

\newcommand{\radiusResult} {\ensuremath{r_0 = 1.249 \pm 0.008\, \mathrm{(stat)} \,^{+0.024} _{-0.021}\, \mathrm{(syst)}}\,fm\xspace}
 
\begin{titlepage}
\PHyear{2019}       
\PHnumber{232}      
\PHdate{25 October}  

\title{Investigation of the p$\mbox{--}\bm{\Sigma}^{\bf{0}}$ interaction via femtoscopy in \pp collisions}
\ShortTitle{Investigation of the \psiZCF interaction via femtoscopy}

\Collaboration{ALICE Collaboration\thanks{See Appendix~\ref{app:collab} for the list of collaboration members}}
\ShortAuthor{ALICE Collaboration} 

\begin{abstract}

This Letter presents the first direct investigation of the \psiZCF interaction, using the femtoscopy technique in high-multiplicity \pp collisions at \thirteen measured by the ALICE detector. The \siZ is reconstructed via the decay channel to $\lmb\phot$, and the subsequent decay of \lmb to p$\pi^-$. The photon is detected via the conversion in material to \ee pairs exploiting the capability of the ALICE detector to measure electrons at low transverse momenta.
The measured \psiZCF correlation indicates a shallow strong interaction. The comparison of the data to several theoretical predictions obtained employing the \textit{Correlation Analysis Tool using the Schr\"odinger Equation} (CATS) and the \Ledn shows that the current experimental precision does not yet allow to discriminate between different models, as it is the case for the available scattering and hypernuclei data. Nevertheless, the \psiZCF correlation function is found to be sensitive to the strong interaction, and driven by the interplay of the different spin and isospin channels.
This pioneering study demonstrates the feasibility of a femtoscopic measurement in the \psiZCF channel and with the expected larger data samples in LHC Run 3 and Run 4, the \psiZCF interaction will be constrained with high precision.

\end{abstract}
\end{titlepage}

\setcounter{page}{2} 

\section{Introduction}
\label{sec:intro}

A quantitative understanding of the hyperon--nucleon interaction in the strangeness $S=-1$ sector is fundamental to pin down the role of strangeness within low energy quantum chromodynamics and to study the properties of baryonic matter at finite densities.
The possible presence of the isoscalar \lmb and the isovector $(\Sigma^+, \, \Sigma^0, \, \Sigma^-)$ hyperon states in the inner core of neutron stars (NS) is currently under debate due to the limited knowledge of the interaction of such hyperons with nuclear matter.
The inclusion of hyperons in the description of the nuclear matter inside NS typically contains only \lmb states, and the on-average attractive nucleon$-$\lmb (\LNCF) interaction leads to rather soft Equations of State (EoS) for NS. These are then unable to provide stability for stars of about two solar masses~\cite{Demorest2SolMass, Antoniadis2SolMass}. The $\Sigma$ hyperons are rarely included in the EoS for NS because of the limited knowledge about the \SNCF strong interaction.

Indeed, while the attractive \LNCF interaction is reasonably well constrained from the available scattering and light hypernuclei data~\cite{LambdapScatt1, LambdapScatt2, LambdaHypernuclei}, the nature of the \SNCF interaction lacks conclusive experimental measurements. 
One of the major complications for experimental studies is the fact that the decay of all \sigm states involves neutral decay products~\cite{PDG}, thus requiring high-resolution calorimeters.

The main source of experimental constraints on the \SNCF system comes from scattering measurements~\cite{EiseleScattS,EngelmannScattS,HeppCapture}, analysis of \siM atoms~\cite{SigmaAtom1, SigmaAtom2, SigmaAtom3}, and hypernuclei production data~\cite{KEK1989He4,BNL1998He4,TangSigma1988BNL,KEK2004}, although the latter are mainly dominated by large statistical uncertainties and large kaon decay background.
Latest hypernuclear results obtained from different nuclear targets point towards an attractive interaction in the isospin $I=1/2$ channel of the \SNCF system~\cite{KEK1989He4, BNL1998He4}, and repulsion in the $I=3/2$ channel~\cite{TangSigma1988BNL, KEK2004}.
Hypernuclear measurements, however, are performed at nuclear saturation density and hence in the presence of more than one nucleon, resulting in a substantial model dependence in the interpretation of the experimental data~\cite{GalRev}.

Additionally, the hyperon--nucleon dynamics are strongly affected by the conversion process \LNCF $\leftrightarrow$ \SNCF, occurring in the  $I=1/2$ channel due to the close kinematic threshold between the two systems (about \SI{80}{MeV})~\cite{JuelichModel, NijmegenModel2010, chiEFT2013,Haidenbauer2019SNcoupling,Haidenbauer2016NS}. 
This coupling is expected to provide an additional attractive contribution in the two-body \LNCF interaction in vacuum~\cite{Haidenbauer2019SNcoupling,Haidenbauer2016NS}.
Indeed, depending on the strength of the \LNCF $\leftrightarrow$ \SNCF coupling at the two-body level, the corresponding in-medium hyperon properties are very different. 
For a strong coupling, this leads to a repulsive single-particle potential $U_\lmb$ at large densities~\cite{Haidenbauer2019SNcoupling,Haidenbauer2016NS}. For the \sigm hyperon, the in-medium properties are mostly determined by the overall repulsion in the $I=3/2$ component~\cite{Haidenbauer2019SNcoupling,Haidenbauer2016NS}.
A repulsive component in the hyperon--nucleon interactions could shift the onset for hyperon production to larger densities, above 2$-$3 times the normal saturation density, thus leading to stiffer EoS which are able to describe the experimental constraint of NS. 

To this end, different theoretical approaches including chiral effective field theories (\chiEFT)~\cite{chiEFT2013} and meson-exchange models with hadronic~\cite{ESC16} and quark~\cite{fss2} degrees of freedom provide a similar description of the available data by assuming a strong repulsion in the spin singlet $S = 0$, $I=1/2$ and spin triplet $S = 1$, $I=3/2$ and an overall attraction in the remaining channels. Recent ab initio lattice calculations at quark physical masses show a similar dependence on spin-isospin configurations for the central potential term~\cite{Lattice}.
The strength of the coupled-channel \LNCF $\leftrightarrow$ \SNCF is strongly model dependent as well. Calculations based on chiral models~\cite{Haidenbauer2019SNcoupling,chiEFT2013} and meson-exchange models~\cite{NSC97,JuelichModel} predict a rather strong or much weaker coupling, respectively.
A self-consistent description of this coupled-channel demands a detailed knowledge of the strong interaction in the \SNCF system.

Recently, the study of two-particle correlations in momentum space measured in ultra-relativistic proton--proton (\pp) and proton--nucleus collisions has proven to provide direct access to the interaction between particle pairs in vacuum~\cite{FemtoRun1, FemtoLambdaLambda, FemtopXi}. The small size of the colliding systems of about \SI{1}{fm} results in a pronounced correlation signal from strong final state interactions, which permits the latter to be precisely constrained. These measurements provided additional data in the hyperon sector with an unprecedented precision in the low momentum regime.
In this Letter, these studies are extended to the \sigm sector. The electromagnetic decay of the \siZ is exploited for the first direct measurement of the \psiZCF interaction in \pp collisions. 
This study paves the way for extending these investigations to the charged \sigm states, in particular in light of the larger data samples expected from the LHC Runs 3 and 4.

\section{Data analysis}
\label{sec:analysis}

This Letter presents results obtained from a data sample of \pp collisions at \thirteen recorded with the ALICE detector~\cite{ALICE, ALICEperf} during the LHC Run 2 (2015--2018). The sample was collected employing a high-multiplicity trigger with the \VZERO detectors, which consist of two small-angle plastic scintillator arrays located on either side of the collision vertex at pseudorapidities $2.8 < \eta < 5.1$ and $-3.7 < \eta < -1.7$~\cite{VZERO}. 
The high-multiplicity trigger is defined by coincident hits in both \VZERO detectors synchronous with the LHC bunch crossing and by additionally requiring the sum of the measured signal amplitudes in the \VZERO to exceed a multiple of the average value in minimum bias collisions. This corresponds, at the analysis level, to the highest multiplicity interval containing the top 0.17\% of all inelastic collisions with at least one charged particle in $|\eta| < 1$ (referred to as INEL > 0).
This data set presents a suitable environment to study correlations due to the enhanced production of strange particles in such collisions~\cite{ALICEstrangeness}. Additionally, the larger charged-particle multiplicity density with respect to the minimum bias sample significantly increases the probability to detect particle pairs. 
The \VZERO is also employed to suppress background events, such as the interaction of beam particles with mechanical structures of the beam line, or beam-gas interactions.
In-bunch pile-up events with more than one collision per bunch crossing are rejected by evaluating the presence of additional event vertices~\cite{ALICEperf}. The remaining contamination from pile-up events is on the percent level and does not influence the final results.

Charged-particle tracking within the ALICE central barrel is conducted with the Inner Tracking System (\ITS)~\cite{ALICE} and the Time Projection Chamber (\TPC)~\cite{TPC}. The detectors are immersed in a homogeneous \SI{0.5}{\tesla} solenoidal magnetic field along the beam direction.
The \ITS consists of six cylindrical layers of high position-resolution silicon detectors placed radially between $3.9$ and \SI{43}{cm} around the beam pipe. The two innermost layers are Silicon Pixel Detectors (\SPD) and cover the pseudorapidity range $|\eta|<2.0$ and $|\eta|<1.4$, respectively. The two intermediate layers are composed of Silicon Drift Detectors, and the two outermost layers are made of double-sided Silicon micro-Strip Detectors (\SSD), covering $|\eta|< 0.9$ and $|\eta|<1.0$, respectively. 
The \TPC consists of a \SI{5}{m} long, cylindrical gaseous detector with full azimuthal coverage in the pseudorapidity range $|\eta| < 0.9$.
Particle identification (PID) is conducted via the measurement of the specific ionization energy loss (\dEdx) with up to 159 reconstructed space points along the particle trajectory.
The Time-Of-Flight (\TOF)~\cite{TOF} detector system is located at a radial distance of \SI{3.7}{m} from the nominal interaction point and consists of Multigap Resistive Plate Chambers covering the full azimuthal angle in $|\eta|<0.9$. PID is accomplished by measuring the particle's velocity $\beta$ via the time of flight of the particles in conjunction with their trajectory. The event collision time is provided as a combination of the measurements in the \TOF and the \TZERO detector, two Cherenkov counter arrays placed at forward rapidity~\cite{TZERO}.

The primary event vertex (PV) is reconstructed with the combined track information of the \ITS and the \TPC, and independently with \SPD tracklets. When both vertex reconstruction methods are available, the difference of the corresponding $z$ coordinates is required to be smaller than \SI{5}{mm}. 
A uniform detector coverage is assured by restricting the maximal deviation between the $z$ coordinate of the reconstructed PV and the nominal interaction point to $\pm$\SI{10}{cm}. A total of \num{1.0e9} high-multiplicity events are used for the analysis after event selection.

The proton candidates are reconstructed following the analysis methods used for minimum bias \pp collisions at \seven~\cite{FemtoRun1} and 13\,TeV~\cite{FemtopXi, FemtoLambdaLambda}, and are selected from the charged-particle tracks reconstructed with the \TPC in the kinematic range $0.5 < \pt < 4.05$\,\GeVc and $|\eta| < 0.8$. The \TPC and \TOF PID capabilities are employed to select proton candidates by the deviation $n_\sigma$ between the signal hypothesis for a proton and the experimental measurement, normalized by the detector resolution $\sigma$. For candidates with $p < 0.75$\,\GeVc, PID is performed with the \TPC only, requiring $|n_\sigma| < 3$. For larger momenta, the PID information of \TPC and \TOF are combined. 
Secondary particles stemming from weak decays or the interaction of primary particles with the detector material contaminate the signal. The corresponding fraction of primary and secondary protons are extracted using Monte Carlo (MC) template fits to the measured distribution of the Distance of Closest Approach (DCA) of the track to the primary vertex~\cite{FemtoRun1}. The MC templates are generated using PYTHIA 8.2~\cite{pythia} and filtered through the ALICE detector~\cite{GEANT} and reconstruction algorithm~\cite{ALICE}. The resulting purity of protons is found to be 99\%, with a primary fraction of 82\%. 

The \siZ is reconstructed via the decay channel $\siZ \rightarrow \lmb \gamma$ with a branching ratio of almost 100\%~\cite{PDG}. The decay is characterized by a short life time rendering the decay products indistinguishable from primary particles produced in the initial collision. Due to the small mass difference between the \lmb and the \siZ of about \SI{77}{MeV/\clight^2}, the \phot has typically energies of only few hundreds of \mev.
Therefore, it is reconstructed relying on conversions to \ee pairs in the detector material of the central barrel exploiting the unique capability of the ALICE detector to identify electrons down to transverse momenta of 0.05\,\GeVc. 
For transverse radii $R < 180$\,cm and $|\eta| < 0.9$ the material budget corresponds to $(11.4 \pm 0.5)$\,\% of a radiation length $X_0$, and accordingly to a conversion probability of $(8.6 \pm 0.4)$\,\%~\cite{PCM2}. Details of the photon conversion analysis and the corresponding selection criteria are described in~\cite{PCM1,PCM2}.
The reconstruction relies on the identification of secondary vertices by forming so-called \Vdecay decay candidates from two oppositely-charged tracks using a procedure described in detail in~\cite{ALICEperfrep}.  The products of the potential $\gamma$ conversion are reconstructed with the \TPC and the \ITS in the kinematic range $\pt > 0.05$\,\GeVc and $|\eta| < 0.9$. The candidates for the \ee pair are identified by a broad PID selection in the \TPC $-6 < n_\sigma < 7$. 
The resulting \phot candidate is obtained as the combination of the daughter tracks. Only candidates with $\pt > 0.02$\,\GeVc and within $|\eta| < 0.9$ are accepted.
Combinatorial background from primary \ee pairs, or Dalitz decays of the $\pi^{0}$ and $\eta$ mesons is removed by requiring that the radial distance of the conversion point, with respect to the detector centre, ranges from \SIrange{5}{180}{cm}. 
Residual contaminations from \kzero and \lmb are removed by a selection in the Armenteros-Podolandski space~\cite{Armenteros, PCM1}.
Random combinations of electrons and positrons are further suppressed by a two-dimensional selection on the angle between the plane defined by the \ee pair, and the magnetic field~\cite{PsiPair} in combination with the reduced $\chi^2$ of a refit of the reconstructed \Vdecay assuming that the particle originates from the primary vertex and has $M_{\Vdecay} = 0$~\cite{PCM1}.
The Cosine of the Pointing Angle (CPA) between the \phot momentum and the vector pointing from the PV to the decay vertex is required to be CPA $ > 0.999$. In addition to the tight CPA selection, the contribution of particles stemming from out-of-bunch pile-up is suppressed by restricting the DCA of the photon to be along the beam direction (DCA$_z < 0.5$\,cm).
After application of the selection criteria, about \num{946e6} \phot candidates with a purity of about 95.4\% are available for further processing. 

The \lmb particle candidates are reconstructed via the subsequent decay $\lmb \rightarrow \mathrm{p}\pi^-$ with a branching ratio of 63.9\%~\cite{PDG}, following the procedures described in~\cite{FemtoRun1, FemtoLambdaLambda}. For the \almb the charge conjugate decay is exploited, and the same selection criteria are applied.
The decay products are reconstructed with the \TPC and the \ITS within $|\eta| < 0.9$. The daughter candidates are identified by a broad PID selection in the TPC $|n_\sigma| < 5$. 
The resulting \lmb candidate is obtained as the combination of the daughter tracks. The contribution of fake candidates is reduced by requesting a minimum $\pt > 0.3$\,\GeVc. The coarse PID selection of the daughter tracks introduces a residual \kzero contamination in the sample of the \lmb candidates.
This contamination is removed by a $1.5\sigma$ rejection on the invariant mass assuming a decay into $\pi^{+}\pi^{-}$, where $\sigma$ corresponds to the width of a Gaussian fitted to the \kzero signal.
Topological selections further enhance the purity of the \lmb sample. The radial distance of the decay vertex with respect to the detector centre ranges from \SIrange{0.2}{100}{cm} and CPA $> 0.999$.
In addition to the tight CPA selection, particles stemming from out-of-bunch pile-up are rejected using the timing information of the \SPD and \SSD, and the \TOF detector. One of the two daughter tracks is required to have a hit in one of these detectors. 
After application of the selection criteria, about \num{188e6} (\num{178e6}) \lmb (\almb) candidates with a purity of 94.6\% (95.3\%) are available for further processing. 

The \siZ (\asiZ) candidates are obtained by combining all \lmb (\almb) and \phot candidates from the same event, where the nominal particle masses~\cite{PDG} are assumed for the daughters. In particular the timing selection on the daughter tracks of the \lmb assures that the \siZ candidates stem from the right bunch crossing.
In case a daughter track is used to construct two \phot, \lmb, and \almb candidates, or a combination thereof, the one with the smaller CPA is removed from the sample. In order to further optimize the yield and the purity of the sample, only \siZ candidates with $\pt > 1$\,\GeVc are used.

\begin{figure*}
    \centering
    \includegraphics[width=0.495 \linewidth]{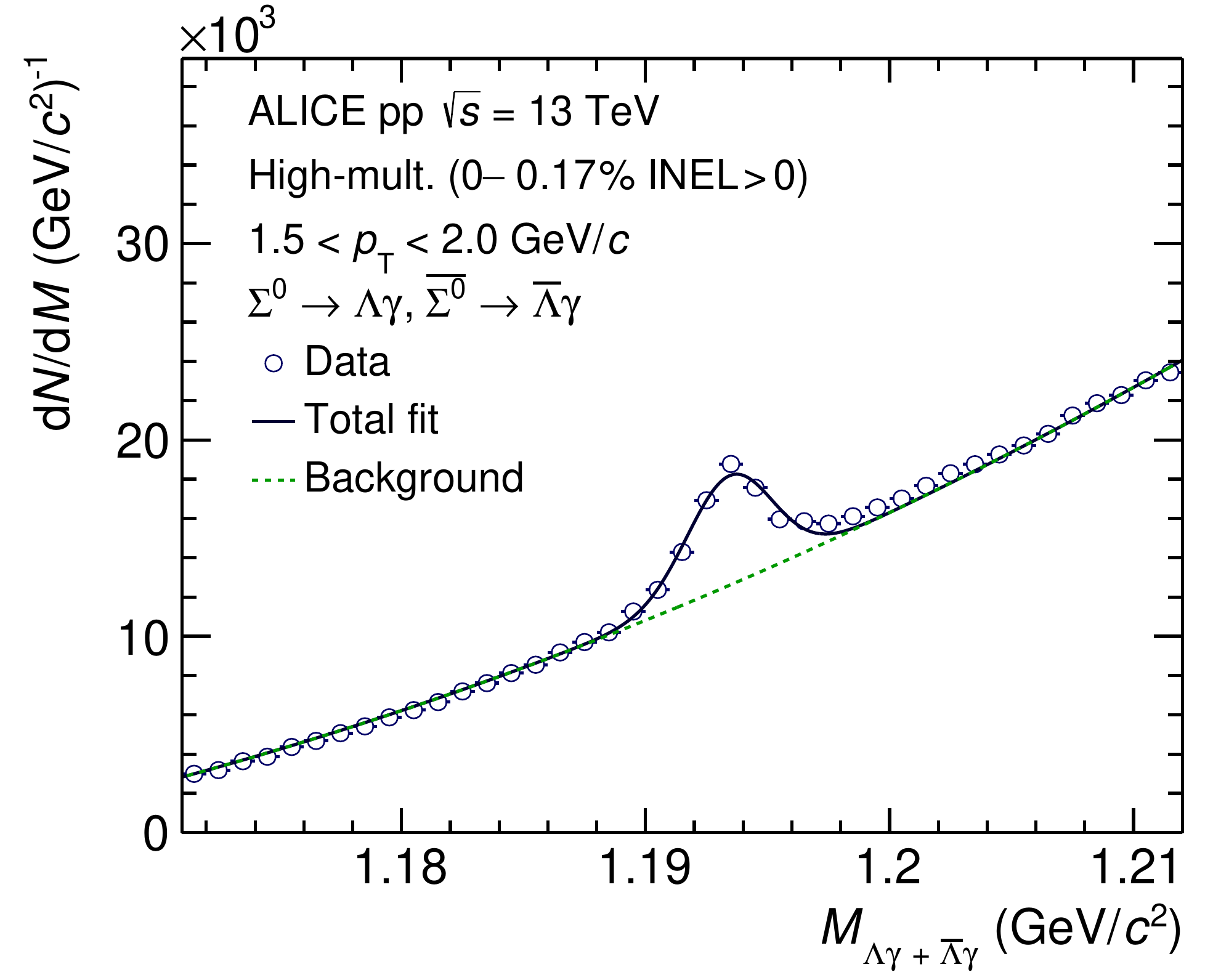}
    \includegraphics[width=0.495 \linewidth]{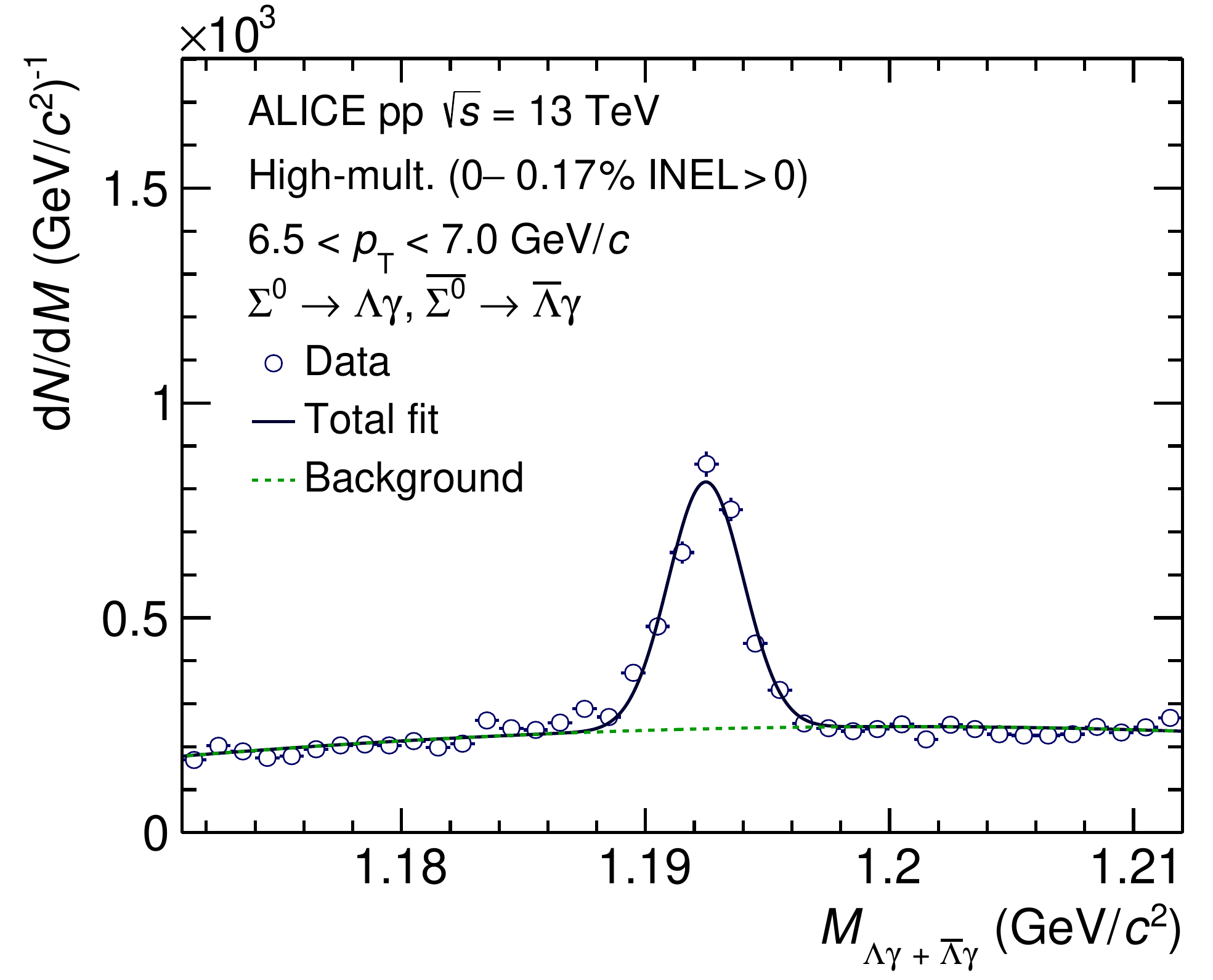}
    \caption{Invariant mass distribution of the $\Lambda\phot$ and $\overline{\Lambda}\phot$ candidates, in two \pt intervals of $1.5 - 2.0$\,\GeVc and $6.5 - 7.0$\,\GeVc. The signal is described by a single Gaussian, and the background by a polynomial of third order. The number of \siZ candidates is evaluated within $M_{\siZ}(\pt) \pm 3$\,\MeVmass. Only statistical uncertainties are shown.}
    \label{fig:SigmaInvMass}
\end{figure*}

The resulting invariant mass spectrum is shown in Fig.~\ref{fig:SigmaInvMass} for two \pt intervals. In order to obtain the raw yield, the signal is fitted with a single Gaussian, and the background with a third-order polynomial. Due to the deteriorating momentum resolution for low \pt tracks, the mean value of the Gaussian $M_{\siZ}$ exhibits a slight \pt dependence, which is well reproduced in MC simulations. The \siZ (\asiZ) candidates for femtoscopy are selected as $M_{\siZ}(\pt) \pm 3$\,\MeVmass. The width of the interval is chosen as a compromise between the candidate counts and purity.
In total, about \num{115e3} (\num{110e3}) \siZ (\asiZ) candidates are found at a purity of about 34.6\%. Due to the enhanced combinatorial background at low \pt, the purity increases from about 20\% at the lower \pt threshold to its saturation value of about 60\% above 5\,\GeVc. 
Only one candidate per event is used, and is randomly selected in the very rare case in which more than one is available. In less than one per mille of the cases when the track of a primary proton is also employed as the daughter track of the \phot or the \lmb, the corresponding \siZ candidate is rejected.
Since only strongly decaying resonances feed to the \siZ~\cite{PDG}, all candidates are considered to be primary particles. 

\section{Analysis of the correlation function}
\label{sec:CF}
The experimental definition of the two-particle correlation function, for both \pP and \psiZCF pairs, is given by~\cite{Lisa:2005dd},
\begin{equation}
C(\kstar)=\mathcal{N}\times\frac{N_{\rm{same}}(\kstar)}{N_{\rm{mixed}}(\kstar)} \xrightarrow{\kstar\rightarrow\infty}1,
\label{eq:CFexp}
\end{equation}
with the same ($N_{\rm{same}}$) and mixed ($N_{\rm{mixed}}$) event distributions of \kstar and a normalization constant $\mathcal{N}$. The relative momentum of the pair \kstar is defined as $\kstar=\frac{1}{2}\times|\mathbf{p}^*_1-\mathbf{p}^*_2|$, where $\mathbf{p}^*_1$ and $\mathbf{p}^*_2$ are the momenta of the two particles in the pair rest frame, denoted by the ${}^*$. 
The normalization is evaluated in $\kstar \in [240, 340]$\,\MeVc for \pP and in $\kstar \in [250, 400]$\,\MeVc for \psiZCF pairs, where effects of final state interactions are absent and hence the correlation function approaches unity.

The trajectories of the \pP and \ApAP pairs at low \kstar are almost collinear, and might therefore be affected by detector effects like track splitting and merging~\cite{Adam:2015vja}. Accordingly, the reconstruction efficiency for pairs in the same and mixed event might differ. To this end, a close-pair rejection criterion is employed removing \pP and \ApAP pairs fulfilling $\sqrt{\Delta\eta^2+\Delta\varphi^{*2}} < 0.01$, where the azimuthal coordinate $\varphi^{*}$ considers the track curvature in the magnetic field.

A total number of \num{1.7e6} (\num{1.3e6}) \pP (\ApAP) and 587 (539) \psiZCF (\apasiZCF) pairs contribute to the respective correlation function in the region $\kstar < 200$\,\MeVc. 
To enhance the statistical significance of the results, the correlation functions of baryon--baryon and antibaryon\mbox{--}antibaryon pairs are combined. Therefore, in the following \psiZCF denotes the combination \psiZCombCF, and correspondingly for \pP. 

The systematic uncertainties of the experimental correlation function are evaluated by simultaneously varying all proton, \lmb, \phot, and \siZ single-particle selection criteria by up to 20\% around the nominal values. Only variations that modify the pair yield by less than 10\% (20\%) for \psiZCF (\pP)  with respect to the default choice are considered, and the \siZ purity by less than 5\%.
The impact of statistical fluctuations is reduced by evaluating the systematic uncertainties in intervals of \SI{100}{MeV/\clight} (\SI{20}{MeV/\clight}) in \kstar for \psiZCF (\pP). The resulting systematic uncertainties are parametrized by an exponential function and interpolated to obtain the final point-by-point uncertainties. At the respectively lowest \kstar, the total systematic uncertainties are of the order of 2.5\% for both \pP and \psiZCF.

Using the femtoscopy formalism~\cite{Lisa:2005dd}, the correlation function can be related to the source function $S(r^*)$ and the two-particle wave function $\Psi(\vec{r^{*}}, \vec{\kstar})$ incorporating the interaction, 
\begin{equation}\label{eq:Ck}
C(\kstar)=\int \mathrm{d}^3 r^{*} ~ S(r^{*}) \mid \Psi(\vec{r^{*}}, \vec{\kstar}) \mid ^2,
\end{equation}
where $r^{*}$ refers to the relative distance between the two particles.
As demonstrated in~\cite{FemtoRun1, FemtoLambdaLambda, FemtopXi, CATS} the correlation function becomes particularly sensitive to the strong interaction for small emission sources formed in \pp and \pPb collisions.
For this study, a spherically symmetric emitting source is assumed, with a Gaussian shaped core density profile parametrized by a radius $r_{0}$, which is obtained from a fit to the \pP correlation function, similarly as in~\cite{FemtoLambdaLambda, FemtopXi}. Following the premise of a common emission source the such extracted radius is then used as an input to fit the \psiZCF correlation function. Possible modifications of the source profile due to the influence of strongly decaying resonances~\cite{StrongDecay1, StrongDecay2, StrongDecay3} are considered in the evaluation of the systematic uncertainties associated with the fitting procedure. 

The genuine \pP correlation function is modeled using the \textit{Correlation Analysis Tool using the Schr\"odinger equation} (CATS)~\cite{CATS}, which allows one to use either a local potential $V(r)$ or directly the two-particle wave function, and additionally any source distribution as input to compute the correlation function.
For the \pP correlation function the strong Argonne $v_{18}$ potential~\cite{Wiringa:1994wb} in the $S$, $P$, and $D$ waves is used as an input to CATS. 

The theoretical correlation function for \psiZCF is modeled employing two different approaches. On the one hand, in CATS the correlation function is computed from the isospin-averaged wave functions obtained within a coupled-channel formalism.
On the other hand, the \Ledn~\cite{Lednicky:1981su} relies on the effective-range expansion using scattering parameters as input to evaluate the correlation function. The coupling of the \nSplusCF system to \psiZCF considering the different thresholds is explicitly included by means of a coupled-channel approach, while the effect of the \pLCF channel is incorporated by complex scattering parameters~\cite{Stavinskiy:2007wb}. \\
Details of the employed models are described in the next Section.

The experimental data are compared with the modeled correlation function considering the finite experimental momentum resolution~\cite{FemtoRun1}. In addition to the genuine correlation function of interest, the measured correlation function also contains residual correlations due to protons coming from weak decays of other particles, such as \lmb and \siP (feed-down), and misidentifications. These effects are included by modeling the total correlation function as a decomposition,
\begin{equation}
    C_{\rm{model}}(\kstar) = 1 + \sum_{i} \lambda_{i} \times (C_{i}(\kstar) - 1),
    \label{eq:CFmodelTotal}
\end{equation}
where the sum runs over all contributions. Their relative contribution is given by the $\lambda$ parameters computed in a data-driven way from single-particle properties such as the purity and feed-down fractions~\cite{FemtoRun1}, and are summarized in Table~\ref{tab:lambdaParam}.

Apart from the genuine \pP correlation function, a significant contribution comes from the decay of \lmb particles feeding to the proton pair. The residual \pLCF correlation function is modeled using either the Usmani potential~\cite{Bodmer:1984gc}, chiral effective field theory calculations at Leading (LO)~\cite{chiEFTLO}, or Next-To-Leading order (NLO)~\cite{chiEFT2013}. The resulting correlation function is transformed into the momentum basis of the \pP pair by applying the corresponding decay matrices~\cite{PhysRevC.89.054916}. All other contributions are assumed to be $C(\kstar)\sim1$.
Due to the challenging reconstruction of the \siZ, the experimental purity of the \siZ sample is rather low, and additionally exhibits a strong dependence on the transverse momentum \pt as demonstrated in Fig.~\ref{fig:SigmaInvMass}. The average \pt of the \siZ candidates used to construct the correlation function at $\kstar < 200$\,\MeVc, however, is lower than the $\langle \pt \rangle$ of all inclusive \siZ candidates. Considering this effect, the \siZ purity employed to compute the $\lambda$ parameters is found to be 27.4\%. 
Accordingly, the main contribution to the \psiZCF correlation function stems from the combinatorial background appearing in the invariant mass spectrum around the \siZ peak, which in the following is referred to as \SB. The shape of the \pSBCF correlation function is extracted from the sidebands of the invariant mass selection, and is found to be independent of the choice of mass window. 
The non-flat behavior is mainly determined by residual \pLCF correlations which are smeared by an uncorrelated \phot, and defines the baseline of the measurement of the \psiZCF correlation function. The shape is parametrized with a Gaussian distribution and weighted by its $\lambda$ parameter. All other contributions stemming from misidentified protons or from feed-down are assumed to be $C(\kstar)\sim1$.

\begin{table}   
\centering
\caption[lambda param]{Weight parameters for the individual components of the measured correlation function. Contributions from feed-down contain the mother particle listed as a sub-index. Non-flat contributions are listed individually.}
\begin{tabular}{l | c | l | c}
\multicolumn{2}{c |}{\pP} & \multicolumn{2}{c}{\psiZCF} \\
Pair & $\lambda$ parameter (\%) & Pair & $\lambda$ parameter (\%)   \\
\hline
\pP & 67.0 & \psiZCF & 22.0 \\
p$_{\Lambda}$\mbox{--p} & 20.3 & \pSBCF & 73.1 \\
Feed-down (flat) & 11.6& Feed-down (flat) & 4.7 \\
Misidentification (flat) & 1.1 & Misidentification (flat) & 0.2 \\
\end{tabular} 
\label{tab:lambdaParam}
\end{table}

\begin{figure}
    \centering
    \includegraphics[width=0.69\linewidth]{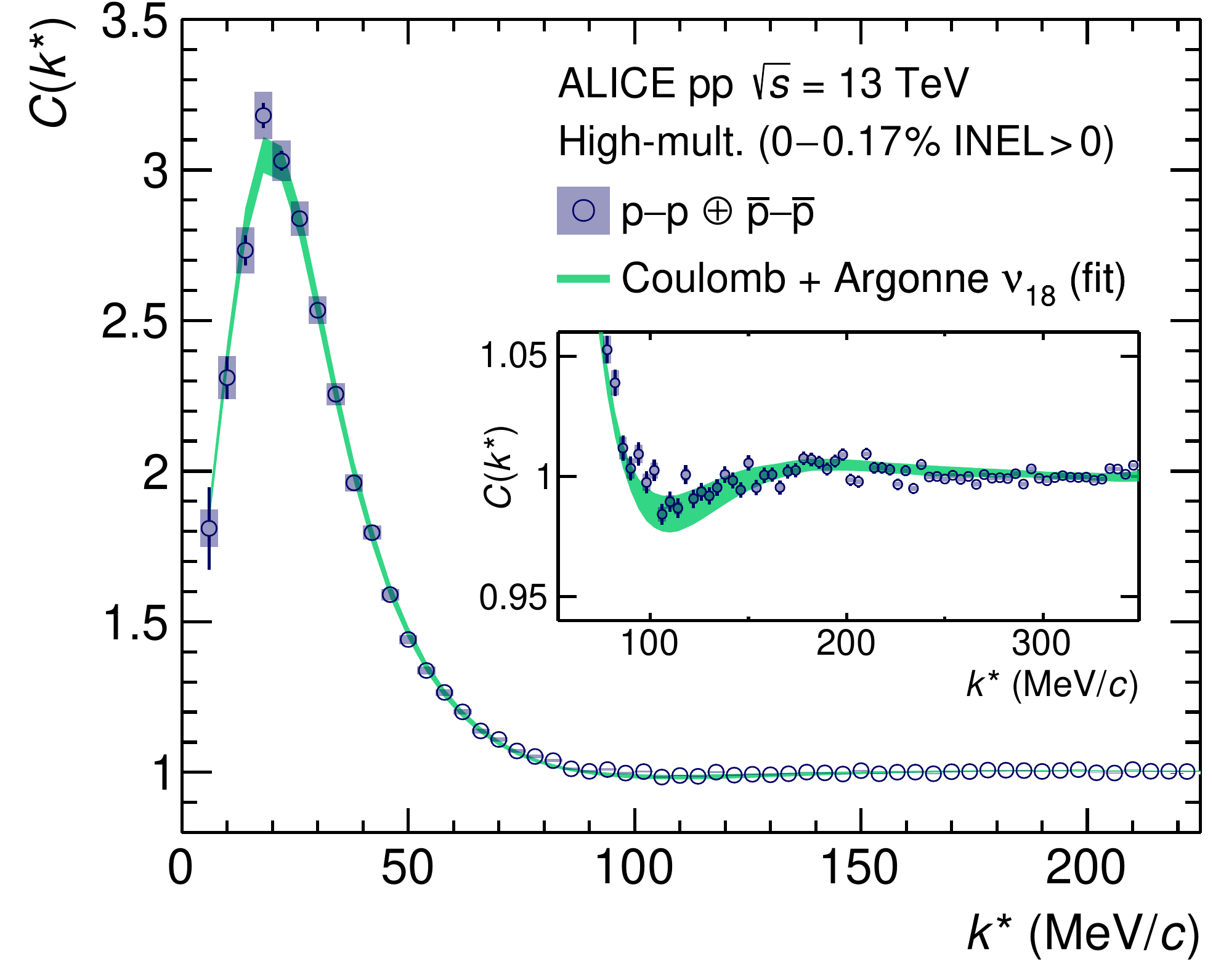}
    \caption{Measured correlation function of \pPComb. Statistical (bars) and systematic uncertainties (boxes) are shown separately. The width of the band corresponds to one standard deviation of the systematic uncertainty of the fit.}
    \label{fig:CFpp}
\end{figure}

The total correlation function including all corrections is then multiplied by a polynomial baseline  $C_{\mathrm{\rm{non-femto}}}(\kstar)$,
\begin{equation}
C(\kstar) = C_{\mathrm{\rm{non-femto}}}(\kstar) \times C_{\rm{model}}(\kstar),
\end{equation}
to account for the normalization and non-femtoscopic background effects stemming e.g.~from momentum and energy conservation~\cite{FemtoRun1}. 
The \pP correlation function is fitted in the range $\kstar\in[\num{0},\num{375}]~\si{MeV\per\clight}$ to determine simultaneously the femtoscopic radius $r_{0}$ and the parameters of the baseline. To assess the systematic uncertainties on $r_0$ related to the fitting procedure the upper limit of the fit region is varied within $\kstar\in[\num{350},\num{400}]~\si{MeV\per\clight}$. The baseline is modeled as a polynomial of zeroth, first, and second order. Additionally, as discussed above, all three models for the \pLCF residual correlation function are employed, and the input to the $\lambda$ parameters is modified by $\pm$20\% while maintaining a constant sum of the primary and secondary fractions.
The \pP correlation function is shown in Fig.~\ref{fig:CFpp}, where the width of the bands corresponds to one standard deviation of the total systematic uncertainty of the fit. 
The inset shows a zoom of the \pP correlation function at intermediate \kstar, where the effect of repulsion becomes apparent. The femtoscopic fit yields a radius of \radiusResult.

Analyses of $\pi$\mbox{--}$\pi$ and K\mbox{--}K correlation functions at ultrarelativistic energies in elementary~\cite{BECLevyCMS} and heavy-ion collisions~\cite{BECLevyPhenix} indicate a source distribution significantly deviating from a Gaussian. Indeed, strongly decaying resonances are known to introduce significant exponential tails to the source distribution, especially for $\pi$\mbox{--}$\pi$ pairs~\cite{StrongDecay1, StrongDecay2, StrongDecay3}. 
This becomes evident when studying the corresponding resonance contributions obtained from the statistical hadronization model within the canonical approach~\cite{Becattini}. The main resonances feeding to pions, $\rho$ and $\omega$, are significantly longer-lived than those feeding to protons ($\Delta$) and \siZ (\lmb{}(1405)).
Hence, it is not surprising that the source distribution for $\pi$\mbox{--}$\pi$ deviates from a Gaussian.
These conclusions are underlined when fitting the \pP correlation function with a L\'evy-stable source distribution~\cite{Levy1, Levy2}.
Leaving both the femtoscopic radius and the stability parameter $\alpha$ for the fit to determine, the Gaussian source shape ($\alpha = 2$) is recovered.
Employing a Cauchy-type source distribution ($\alpha = 1$), the data cannot be satisfactorily described. Therefore, the premise of a Gaussian source holds for baryon--baryon pairs.

Accordingly, a Gaussian source with femtoscopic radius $r_0$ is used to fit the \psiZCF correlation function.
The parameters of the linear baseline are obtained from a fit to the \pSBCF correlation function in $\kstar\in$~[\num{250},\,\num{600}]~\si{MeV\per\clight}, where it is consistent and kinematically comparable with \psiZCF, however featuring significantly smaller uncertainties.
The experimental \psiZCF correlation function is then fitted in the range $\kstar < 550$\,\MeVc, and varied during the fitting procedure within $\kstar \in [500, 600]$\,\MeVc to determine the systematic uncertainty. Additionally, the input to the $\lambda$ parameters is modified by $\pm$20\% while maintaining a constant sum of the primary and secondary fractions.
The parameters of the baseline are varied within $1\sigma$ of their uncertainties considering their correlation, including the case of a constant baseline. 
Finally, the femtoscopic radius is varied according to its uncertainties.
Possible variations of the \psiZCF source due to contributions of \mt scaling and strong decays are incorporated by decreasing $r_0$ by 15\%, similarly as in~\cite{FemtopXi, FemtoLambdaLambda}. The corresponding resonance yields are taken from the statistical hadronization model within the canonical approach~\cite{Becattini}. \\
All correlation functions resulting from the above mentioned variations of the selection criteria are fitted during the procedure, additionally considering variations of the mass window to extract the \pSBCF baseline. 
The width of the bands in Fig.~\ref{fig:CFpSigma0} corresponds to one standard deviation of the total systematic uncertainty of the fit. The absolute correlated uncertainty due to the modeling of the \pSBCF baseline correlation function is shown separately at the bottom of the figure.

\section{Results}
\label{sec:results}
\begin{figure}
    \centering
    \includegraphics[width=0.69\linewidth]{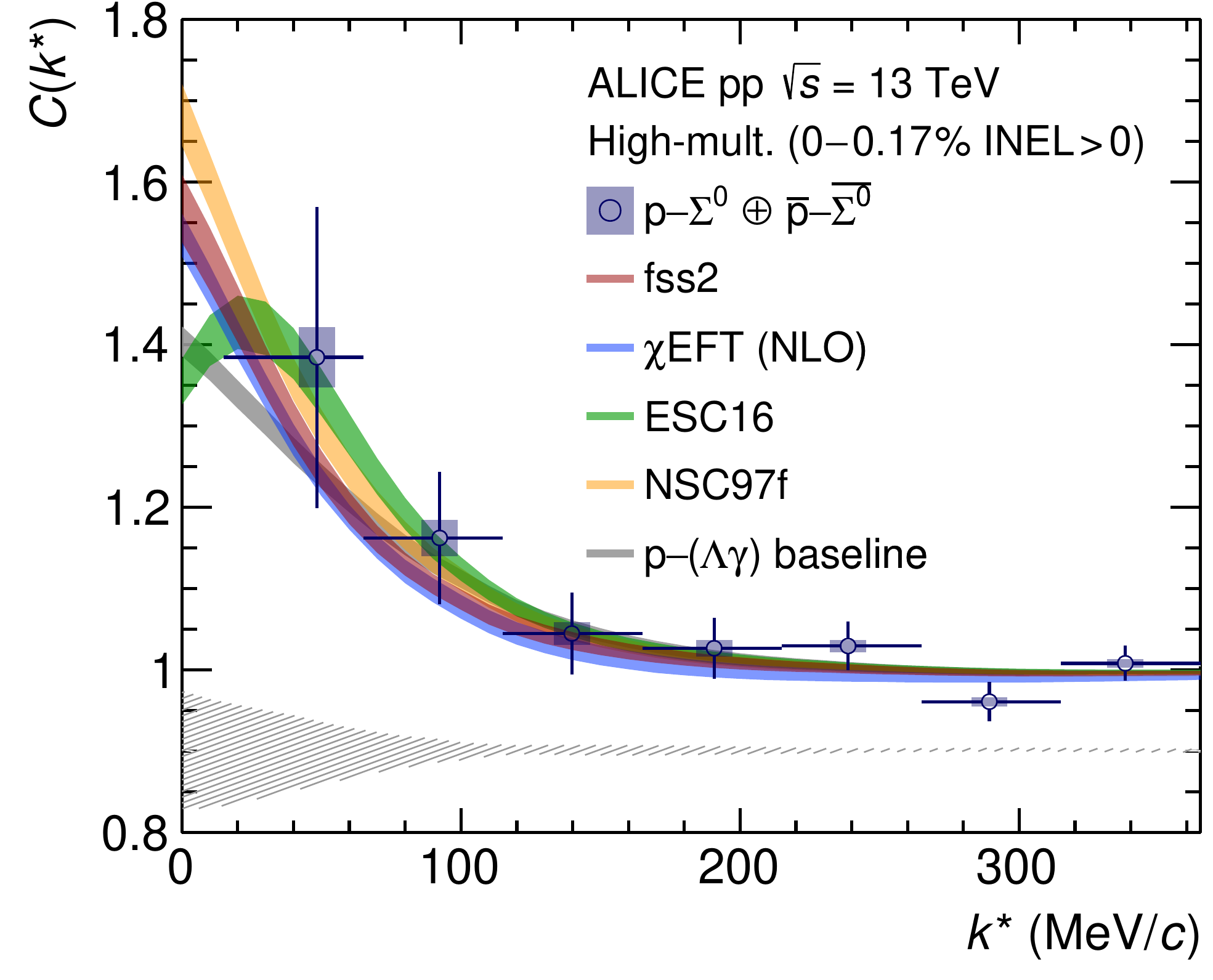}
    \caption{Measured correlation function of \psiZCombCF. Statistical (bars) and systematic uncertainties (boxes) are shown separately. The gray band denotes the \pSBCF baseline. The data are compared with different theoretical models. The corresponding correlation functions are computed using CATS~\cite{CATS} for \chiEFT~\cite{chiEFT2013}, \NSC~\cite{NSC97} and \ESC~\cite{ESC16}, and using the \Ledn~\cite{Lednicky:1981su, Stavinskiy:2007wb} for \fss~\cite{fss2}. The width of the bands corresponds to one standard deviation of the systematic uncertainty of the fit. The absolute correlated uncertainty due to the modeling of the \pSBCF baseline is shown separately as the hatched area at the bottom of the figure.}
    \label{fig:CFpSigma0}
\end{figure}

The experimental \psiZCombCF correlation function is shown in Fig.~\ref{fig:CFpSigma0}. 
The \kstar value of the data points is chosen according to the $\langle$\kstar$\rangle$ of the same event distribution $N_{\rm{same}}(\kstar)$ in the corresponding interval. Therefore, due to the low number of counts in the first bin, the data point is shifted with respect to the bin center.
Since the uncertainties of the data are sizable, a direct determination of scattering parameters via a femtoscopic fit is not feasible. Instead, the data are directly compared with the various models of the interaction.
These include, on the one hand, meson-exchange models, such as \fss~\cite{fss2} and two versions of soft-core Nijmegen models (\ESC~\cite{ESC16}, \NSC ~\cite{NSC97f}), and on the other hand results of \chiEFT at Next-to-Leading Order (NLO)~\cite{chiEFT2013}. The correlation function is modeled using the \Ledn~\cite{Lednicky:1981su} considering the couplings of the \psiZCF system to \pLCF and \nSplusCF~\cite{Stavinskiy:2007wb} with scattering parameters extracted from the \fss model. For the case of \ESC, \NSC and \chiEFT, the wave function of the \psiZCF system, including the couplings, is used as an input to CATS to compute the correlation function. The degree of consistency of the data with the discussed models is expressed by the number of standard deviations $n_{\sigma}$, computed in the range $\kstar < 150$\,\MeVc from the p-value of the theoretical curves. The range of $n_{\sigma}$ shown in Table~\ref{tab:nSigmaTable} is computed as one standard deviation of the corresponding distribution. 
The data are within (0.2$-$0.8)$\sigma$ consistent with the \pSBCF baseline, indicating the presence of an overall shallow strong potential in the \psiZCF channel. 
The main source of uncertainty of the modeling of the correlation function is the parametrization of the \pSBCF baseline due the sizeable statistical uncertainties of the latter.

All employed models for the \SNCF interaction potential succeed in reproducing the scattering data in the $S=-1$ sector~\cite{EiseleScattS}. 
Due to the available experimental constraints, the overall description of the \pLCF interaction yields a consistent description. 
On the other hand, the corresponding \psiZCF correlation functions differ significantly among each other.
This demonstrates that femtoscopic measurements can discriminate and constrain models, and therefore represent a unique probe to study the \SNCF interaction. 
Both \fss and \chiEFT exhibit an overall repulsion in \SNCF at intermediate \kstar, which mainly occurs in the spin singlet $S = 0$, $I=1/2$ and spin triplet $S = 1$, $I=3/2$  components~\cite{fss2,chiEFT2013}. In the low momentum region, below roughly $50$\,\MeVc, both models yield attraction, which is reflected in the profile of the correlation function. 
The Nijmegen models, on the other hand, are characterized by a rather constant attraction over the whole range of \kstar. In particular at low relative momenta, however, the behavior of the two models deviates significantly. 
The shape of the correlation function of the most recent Nijmegen model, \ESC, differs significantly from that of the other calculations. This is mainly due to the fact that the occurrence of bound states in the strangeness sector ($S= -1,-2,-3$) is not allowed in the model~\cite{ESC16}. This leads to a repulsive core in all the \SNCF channels, which can well be observed in Fig.~\ref{fig:CFpSigma0} as the non-monotonic behavior at small relative momenta.
In contrast to all other discussed models, \NSC yields attraction in the spin triplet $S = 1$, $I=3/2$ channel~\cite{NSC97f}. Accordingly, the corresponding correlation function demonstrates the strongest attraction at low momenta.
The rather large differences among the modeled \psiZCF correlation functions demonstrate that the shape of the latter is very sensitive to details of the strong interaction, and driven by the interplay of the different spin and isospin channels. 
This shows the strength of femtoscopic measurements, in particular in the \SNCF channel.

\begin{table*}   
\centering
\caption{Degree of consistency of the different models with the experimental correlation function.}
\begin{tabular}{l | c | c | c | c | c }
Model &  \pSBCF baseline & fss2 & \chiEFT & \NSC & \ESC   \\
\hline
$n_{\sigma}$ $(\kstar < 150$\,\MeVc) & 0.2$-$0.8 & 0.2$-$0.9 & 0.3$-$1.0 & 0.2$-$0.6 & 0.1$-$0.5 \\
\end{tabular} 
\label{tab:nSigmaTable}
\end{table*}

The underlying two-body \SNCF interaction obtained within these models, however, translates into significantly different values for the in-medium single-particle potential $U_{\sigm}$ when included in many-body calculations.
Both the \fss quark-model, along with \chiEFT, deliver similar results at nuclear saturation density, leading to an overall repulsive $U_{\Sigma}$ of around 10$-$17\,MeV~\cite{fss2,Haidenbauer2019SNcoupling,chiEFT2013}. This is in agreement with evidence from relativistic mean field calculations fitting experimental data of \siM atoms~\cite{SigmaAtom3} and the experimental absence of bound states in \sigm hypernuclei~\cite{KEK2004}. On the contrary, both Nijmegen models yield a slightly attractive \sigm single-particle potential, ranging from $\approx -16$\,MeV for \NSC to $\approx -3$\,MeV for \ESC.
As already mentioned, however, the interpretation of hypernuclear measurements introduces a significant model dependence. This concerns not only the extraction of the experimental results, relying for instance on the framework of the distorted-wave impulse approximation~\cite{GalRev}, but also the extrapolation of theoretical calculations to finite density via e.g.~the G-matrix approach~\cite{Gmatrix1,Gmatrix2}. 

\section{Summary}
\label{sec:summary}
This Letter presents the first direct investigation of the \psiZCF interaction in high-multiplicity \pp collisions at \thirteen, hence proving the feasibility of femtoscopic studies in the \SNCF sector.
The \psiZCF correlation function is consistent with the \pSBCF baseline, and therefore the measurement indicates the presence of an overall shallow strong potential.
The data are compared with state-of-the-art descriptions of the interaction, including chiral effective field theory and meson-exchange models. Due to the scarce experimental constraints in the \SNCF sector, the modeled correlation functions differ significantly among each other.
The shape of the modeled correlation functions appears to be very sensitive to details of the strong interaction, and driven by the interplay of the different spin and isospin channels. This proves that femtoscopic measurements in high-energy \pp collisions provide a direct study of the genuine two-body \SNCF strong interaction. 
The presented femtoscopic data cannot discriminate between different models, which is also the case for the available scattering and hypernuclei data. \\
Further femtoscopic studies enabled by the about two orders of magnitude larger \pp data samples of 6\,pb$^{-1}$ in minimum bias collisions at \fivefive and of 200\,pb$^{-1}$ in high-multiplicity at \forteen, foreseen to be collected in the LHC Runs 3 and 4~\cite{ALICE-Run3}, will therefore shed light on the \SNCF sector and provide constraints on the models describing the interaction.

\newenvironment{acknowledgement}{\relax}{\relax}
\begin{acknowledgement}
\section*{Acknowledgements}
The ALICE Collaboration is grateful to J.~Haidenbauer and T.~Rijken for valuable discussions and for providing the theoretical results for the \psiZCF interaction.


The ALICE Collaboration would like to thank all its engineers and technicians for their invaluable contributions to the construction of the experiment and the CERN accelerator teams for the outstanding performance of the LHC complex.
The ALICE Collaboration gratefully acknowledges the resources and support provided by all Grid centres and the Worldwide LHC Computing Grid (WLCG) collaboration.
The ALICE Collaboration acknowledges the following funding agencies for their support in building and running the ALICE detector:
A. I. Alikhanyan National Science Laboratory (Yerevan Physics Institute) Foundation (ANSL), State Committee of Science and World Federation of Scientists (WFS), Armenia;
Austrian Academy of Sciences, Austrian Science Fund (FWF): [M 2467-N36] and Nationalstiftung f\"{u}r Forschung, Technologie und Entwicklung, Austria;
Ministry of Communications and High Technologies, National Nuclear Research Center, Azerbaijan;
Conselho Nacional de Desenvolvimento Cient\'{\i}fico e Tecnol\'{o}gico (CNPq), Financiadora de Estudos e Projetos (Finep), Funda\c{c}\~{a}o de Amparo \`{a} Pesquisa do Estado de S\~{a}o Paulo (FAPESP) and Universidade Federal do Rio Grande do Sul (UFRGS), Brazil;
Ministry of Education of China (MOEC) , Ministry of Science \& Technology of China (MSTC) and National Natural Science Foundation of China (NSFC), China;
Ministry of Science and Education and Croatian Science Foundation, Croatia;
Centro de Aplicaciones Tecnol\'{o}gicas y Desarrollo Nuclear (CEADEN), Cubaenerg\'{\i}a, Cuba;
Ministry of Education, Youth and Sports of the Czech Republic, Czech Republic;
The Danish Council for Independent Research | Natural Sciences, the VILLUM FONDEN and Danish National Research Foundation (DNRF), Denmark;
Helsinki Institute of Physics (HIP), Finland;
Commissariat \`{a} l'Energie Atomique (CEA), Institut National de Physique Nucl\'{e}aire et de Physique des Particules (IN2P3) and Centre National de la Recherche Scientifique (CNRS) and R\'{e}gion des  Pays de la Loire, France;
Bundesministerium f\"{u}r Bildung und Forschung (BMBF) and GSI Helmholtzzentrum f\"{u}r Schwerionenforschung GmbH, Germany;
General Secretariat for Research and Technology, Ministry of Education, Research and Religions, Greece;
National Research, Development and Innovation Office, Hungary;
Department of Atomic Energy Government of India (DAE), Department of Science and Technology, Government of India (DST), University Grants Commission, Government of India (UGC) and Council of Scientific and Industrial Research (CSIR), India;
Indonesian Institute of Science, Indonesia;
Centro Fermi - Museo Storico della Fisica e Centro Studi e Ricerche Enrico Fermi and Istituto Nazionale di Fisica Nucleare (INFN), Italy;
Institute for Innovative Science and Technology , Nagasaki Institute of Applied Science (IIST), Japanese Ministry of Education, Culture, Sports, Science and Technology (MEXT) and Japan Society for the Promotion of Science (JSPS) KAKENHI, Japan;
Consejo Nacional de Ciencia (CONACYT) y Tecnolog\'{i}a, through Fondo de Cooperaci\'{o}n Internacional en Ciencia y Tecnolog\'{i}a (FONCICYT) and Direcci\'{o}n General de Asuntos del Personal Academico (DGAPA), Mexico;
Nederlandse Organisatie voor Wetenschappelijk Onderzoek (NWO), Netherlands;
The Research Council of Norway, Norway;
Commission on Science and Technology for Sustainable Development in the South (COMSATS), Pakistan;
Pontificia Universidad Cat\'{o}lica del Per\'{u}, Peru;
Ministry of Science and Higher Education and National Science Centre, Poland;
Korea Institute of Science and Technology Information and National Research Foundation of Korea (NRF), Republic of Korea;
Ministry of Education and Scientific Research, Institute of Atomic Physics and Ministry of Research and Innovation and Institute of Atomic Physics, Romania;
Joint Institute for Nuclear Research (JINR), Ministry of Education and Science of the Russian Federation, National Research Centre Kurchatov Institute, Russian Science Foundation and Russian Foundation for Basic Research, Russia;
Ministry of Education, Science, Research and Sport of the Slovak Republic, Slovakia;
National Research Foundation of South Africa, South Africa;
Swedish Research Council (VR) and Knut \& Alice Wallenberg Foundation (KAW), Sweden;
European Organization for Nuclear Research, Switzerland;
Suranaree University of Technology (SUT), National Science and Technology Development Agency (NSDTA) and Office of the Higher Education Commission under NRU project of Thailand, Thailand;
Turkish Atomic Energy Agency (TAEK), Turkey;
National Academy of  Sciences of Ukraine, Ukraine;
Science and Technology Facilities Council (STFC), United Kingdom;
National Science Foundation of the United States of America (NSF) and United States Department of Energy, Office of Nuclear Physics (DOE NP), United States of America.
\end{acknowledgement}

\bibliographystyle{utphys}   
\bibliography{bibliography}

\newpage
\appendix

%
%

\section{The ALICE Collaboration}
\label{app:collab}

\begingroup
\small
\begin{flushleft}
S.~Acharya\Irefn{org141}\And 
D.~Adamov\'{a}\Irefn{org94}\And 
A.~Adler\Irefn{org74}\And 
J.~Adolfsson\Irefn{org80}\And 
M.M.~Aggarwal\Irefn{org99}\And 
G.~Aglieri Rinella\Irefn{org33}\And 
M.~Agnello\Irefn{org30}\And 
N.~Agrawal\Irefn{org10}\textsuperscript{,}\Irefn{org53}\And 
Z.~Ahammed\Irefn{org141}\And 
S.~Ahmad\Irefn{org16}
% ...
\And
\renewcommand\labelenumi{\textsuperscript{\theenumi}~}

\section*{Affiliation notes}
\renewcommand\theenumi{\roman{enumi}}
\begin{Authlist}
\item \Adef{org*}Deceased
\item \Adef{orgI}Dipartimento DET del Politecnico di Torino, Turin, Italy
\item \Adef{orgII}M.V. Lomonosov Moscow State University, D.V. Skobeltsyn Institute of Nuclear, Physics, Moscow, Russia
\item \Adef{orgIII}Department of Applied Physics, Aligarh Muslim University, Aligarh, India
\item \Adef{orgIV}Institute of Theoretical Physics, University of Wroclaw, Poland
\end{Authlist}

\section*{Collaboration Institutes}
\renewcommand\theenumi{\arabic{enumi}~}
\begin{Authlist}
\item \Idef{org1}A.I. Alikhanyan National Science Laboratory (Yerevan Physics Institute) Foundation, Yerevan, Armenia
\item \Idef{org2}Bogolyubov Institute for Theoretical Physics, National Academy of Sciences of Ukraine, Kiev, Ukraine
\item \Idef{org3}Bose Institute, Department of Physics  and Centre for Astroparticle Physics and Space Science (CAPSS), Kolkata, India
\item \Idef{org4}Budker Institute for Nuclear Physics, Novosibirsk, Russia
\item \Idef{org5}California Polytechnic State University, San Luis Obispo, California, United States
\item \Idef{org6}Central China Normal University, Wuhan, China
\item \Idef{org7}Centre de Calcul de l'IN2P3, Villeurbanne, Lyon, France
\item \Idef{org8}Centro de Aplicaciones Tecnol\'{o}gicas y Desarrollo Nuclear (CEADEN), Havana, Cuba
\item \Idef{org9}Centro de Investigaci\'{o}n y de Estudios Avanzados (CINVESTAV), Mexico City and M\'{e}rida, Mexico
\item \Idef{org10}Centro Fermi - Museo Storico della Fisica e Centro Studi e Ricerche ``Enrico Fermi', Rome, Italy
\item \Idef{org11}Chicago State University, Chicago, Illinois, United States
\item \Idef{org12}China Institute of Atomic Energy, Beijing, China
\item \Idef{org13}Comenius University Bratislava, Faculty of Mathematics, Physics and Informatics, Bratislava, Slovakia
\item \Idef{org14}COMSATS University Islamabad, Islamabad, Pakistan
\item \Idef{org15}Creighton University, Omaha, Nebraska, United States
\item \Idef{org16}Department of Physics, Aligarh Muslim University, Aligarh, India
\item \Idef{org17}Department of Physics, Pusan National University, Pusan, Republic of Korea
\item \Idef{org18}Department of Physics, Sejong University, Seoul, Republic of Korea
\item \Idef{org19}Department of Physics, University of California, Berkeley, California, United States
\item \Idef{org20}Department of Physics, University of Oslo, Oslo, Norway
\item \Idef{org21}Department of Physics and Technology, University of Bergen, Bergen, Norway
\item \Idef{org22}Dipartimento di Fisica dell'Universit\`{a} 'La Sapienza' and Sezione INFN, Rome, Italy
\item \Idef{org23}Dipartimento di Fisica dell'Universit\`{a} and Sezione INFN, Cagliari, Italy
\item \Idef{org24}Dipartimento di Fisica dell'Universit\`{a} and Sezione INFN, Trieste, Italy
\item \Idef{org25}Dipartimento di Fisica dell'Universit\`{a} and Sezione INFN, Turin, Italy
\item \Idef{org26}Dipartimento di Fisica e Astronomia dell'Universit\`{a} and Sezione INFN, Bologna, Italy
\item \Idef{org27}Dipartimento di Fisica e Astronomia dell'Universit\`{a} and Sezione INFN, Catania, Italy
\item \Idef{org28}Dipartimento di Fisica e Astronomia dell'Universit\`{a} and Sezione INFN, Padova, Italy
\item \Idef{org29}Dipartimento di Fisica `E.R.~Caianiello' dell'Universit\`{a} and Gruppo Collegato INFN, Salerno, Italy
\item \Idef{org30}Dipartimento DISAT del Politecnico and Sezione INFN, Turin, Italy
\item \Idef{org31}Dipartimento di Scienze e Innovazione Tecnologica dell'Universit\`{a} del Piemonte Orientale and INFN Sezione di Torino, Alessandria, Italy
\item \Idef{org32}Dipartimento Interateneo di Fisica `M.~Merlin' and Sezione INFN, Bari, Italy
\item \Idef{org33}European Organization for Nuclear Research (CERN), Geneva, Switzerland
\item \Idef{org34}Faculty of Electrical Engineering, Mechanical Engineering and Naval Architecture, University of Split, Split, Croatia
\item \Idef{org35}Faculty of Engineering and Science, Western Norway University of Applied Sciences, Bergen, Norway
\item \Idef{org36}Faculty of Nuclear Sciences and Physical Engineering, Czech Technical University in Prague, Prague, Czech Republic
\item \Idef{org37}Faculty of Science, P.J.~\v{S}af\'{a}rik University, Ko\v{s}ice, Slovakia
\item \Idef{org38}Frankfurt Institute for Advanced Studies, Johann Wolfgang Goethe-Universit\"{a}t Frankfurt, Frankfurt, Germany
\item \Idef{org39}Fudan University, Shanghai, China
\item \Idef{org40}Gangneung-Wonju National University, Gangneung, Republic of Korea
\item \Idef{org41}Gauhati University, Department of Physics, Guwahati, India
\item \Idef{org42}Helmholtz-Institut f\"{u}r Strahlen- und Kernphysik, Rheinische Friedrich-Wilhelms-Universit\"{a}t Bonn, Bonn, Germany
\item \Idef{org43}Helsinki Institute of Physics (HIP), Helsinki, Finland
\item \Idef{org44}High Energy Physics Group,  Universidad Aut\'{o}noma de Puebla, Puebla, Mexico
\item \Idef{org45}Hiroshima University, Hiroshima, Japan
\item \Idef{org46}Hochschule Worms, Zentrum  f\"{u}r Technologietransfer und Telekommunikation (ZTT), Worms, Germany
\item \Idef{org47}Horia Hulubei National Institute of Physics and Nuclear Engineering, Bucharest, Romania
\item \Idef{org48}Indian Institute of Technology Bombay (IIT), Mumbai, India
\item \Idef{org49}Indian Institute of Technology Indore, Indore, India
\item \Idef{org50}Indonesian Institute of Sciences, Jakarta, Indonesia
\item \Idef{org51}INFN, Laboratori Nazionali di Frascati, Frascati, Italy
\item \Idef{org52}INFN, Sezione di Bari, Bari, Italy
\item \Idef{org53}INFN, Sezione di Bologna, Bologna, Italy
\item \Idef{org54}INFN, Sezione di Cagliari, Cagliari, Italy
\item \Idef{org55}INFN, Sezione di Catania, Catania, Italy
\item \Idef{org56}INFN, Sezione di Padova, Padova, Italy
\item \Idef{org57}INFN, Sezione di Roma, Rome, Italy
\item \Idef{org58}INFN, Sezione di Torino, Turin, Italy
\item \Idef{org59}INFN, Sezione di Trieste, Trieste, Italy
\item \Idef{org60}Inha University, Incheon, Republic of Korea
\item \Idef{org61}Institut de Physique Nucl\'{e}aire d'Orsay (IPNO), Institut National de Physique Nucl\'{e}aire et de Physique des Particules (IN2P3/CNRS), Universit\'{e} de Paris-Sud, Universit\'{e} Paris-Saclay, Orsay, France
\item \Idef{org62}Institute for Nuclear Research, Academy of Sciences, Moscow, Russia
\item \Idef{org63}Institute for Subatomic Physics, Utrecht University/Nikhef, Utrecht, Netherlands
\item \Idef{org64}Institute of Experimental Physics, Slovak Academy of Sciences, Ko\v{s}ice, Slovakia
\item \Idef{org65}Institute of Physics, Homi Bhabha National Institute, Bhubaneswar, India
\item \Idef{org66}Institute of Physics of the Czech Academy of Sciences, Prague, Czech Republic
\item \Idef{org67}Institute of Space Science (ISS), Bucharest, Romania
\item \Idef{org68}Institut f\"{u}r Kernphysik, Johann Wolfgang Goethe-Universit\"{a}t Frankfurt, Frankfurt, Germany
\item \Idef{org69}Instituto de Ciencias Nucleares, Universidad Nacional Aut\'{o}noma de M\'{e}xico, Mexico City, Mexico
\item \Idef{org70}Instituto de F\'{i}sica, Universidade Federal do Rio Grande do Sul (UFRGS), Porto Alegre, Brazil
\item \Idef{org71}Instituto de F\'{\i}sica, Universidad Nacional Aut\'{o}noma de M\'{e}xico, Mexico City, Mexico
\item \Idef{org72}iThemba LABS, National Research Foundation, Somerset West, South Africa
\item \Idef{org73}Jeonbuk National University, Jeonju, Republic of Korea
\item \Idef{org74}Johann-Wolfgang-Goethe Universit\"{a}t Frankfurt Institut f\"{u}r Informatik, Fachbereich Informatik und Mathematik, Frankfurt, Germany
\item \Idef{org75}Joint Institute for Nuclear Research (JINR), Dubna, Russia
\item \Idef{org76}Korea Institute of Science and Technology Information, Daejeon, Republic of Korea
\item \Idef{org77}KTO Karatay University, Konya, Turkey
\item \Idef{org78}Laboratoire de Physique Subatomique et de Cosmologie, Universit\'{e} Grenoble-Alpes, CNRS-IN2P3, Grenoble, France
\item \Idef{org79}Lawrence Berkeley National Laboratory, Berkeley, California, United States
\item \Idef{org80}Lund University Department of Physics, Division of Particle Physics, Lund, Sweden
\item \Idef{org81}Nagasaki Institute of Applied Science, Nagasaki, Japan
\item \Idef{org82}Nara Women{'}s University (NWU), Nara, Japan
\item \Idef{org83}National and Kapodistrian University of Athens, School of Science, Department of Physics , Athens, Greece
\item \Idef{org84}National Centre for Nuclear Research, Warsaw, Poland
\item \Idef{org85}National Institute of Science Education and Research, Homi Bhabha National Institute, Jatni, India
\item \Idef{org86}National Nuclear Research Center, Baku, Azerbaijan
\item \Idef{org87}National Research Centre Kurchatov Institute, Moscow, Russia
\item \Idef{org88}Niels Bohr Institute, University of Copenhagen, Copenhagen, Denmark
\item \Idef{org89}Nikhef, National institute for subatomic physics, Amsterdam, Netherlands
\item \Idef{org90}NRC Kurchatov Institute IHEP, Protvino, Russia
\item \Idef{org91}NRC «Kurchatov Institute»  - ITEP, Moscow, Russia
\item \Idef{org92}NRNU Moscow Engineering Physics Institute, Moscow, Russia
\item \Idef{org93}Nuclear Physics Group, STFC Daresbury Laboratory, Daresbury, United Kingdom
\item \Idef{org94}Nuclear Physics Institute of the Czech Academy of Sciences, \v{R}e\v{z} u Prahy, Czech Republic
\item \Idef{org95}Oak Ridge National Laboratory, Oak Ridge, Tennessee, United States
\item \Idef{org96}Ohio State University, Columbus, Ohio, United States
\item \Idef{org97}Petersburg Nuclear Physics Institute, Gatchina, Russia
\item \Idef{org98}Physics department, Faculty of science, University of Zagreb, Zagreb, Croatia
\item \Idef{org99}Physics Department, Panjab University, Chandigarh, India
\item \Idef{org100}Physics Department, University of Jammu, Jammu, India
\item \Idef{org101}Physics Department, University of Rajasthan, Jaipur, India
\item \Idef{org102}Physikalisches Institut, Eberhard-Karls-Universit\"{a}t T\"{u}bingen, T\"{u}bingen, Germany
\item \Idef{org103}Physikalisches Institut, Ruprecht-Karls-Universit\"{a}t Heidelberg, Heidelberg, Germany
\item \Idef{org104}Physik Department, Technische Universit\"{a}t M\"{u}nchen, Munich, Germany
\item \Idef{org105}Politecnico di Bari, Bari, Italy
\item \Idef{org106}Research Division and ExtreMe Matter Institute EMMI, GSI Helmholtzzentrum f\"ur Schwerionenforschung GmbH, Darmstadt, Germany
\item \Idef{org107}Rudjer Bo\v{s}kovi\'{c} Institute, Zagreb, Croatia
\item \Idef{org108}Russian Federal Nuclear Center (VNIIEF), Sarov, Russia
\item \Idef{org109}Saha Institute of Nuclear Physics, Homi Bhabha National Institute, Kolkata, India
\item \Idef{org110}School of Physics and Astronomy, University of Birmingham, Birmingham, United Kingdom
\item \Idef{org111}Secci\'{o}n F\'{\i}sica, Departamento de Ciencias, Pontificia Universidad Cat\'{o}lica del Per\'{u}, Lima, Peru
\item \Idef{org112}St. Petersburg State University, St. Petersburg, Russia
\item \Idef{org113}Stefan Meyer Institut f\"{u}r Subatomare Physik (SMI), Vienna, Austria
\item \Idef{org114}SUBATECH, IMT Atlantique, Universit\'{e} de Nantes, CNRS-IN2P3, Nantes, France
\item \Idef{org115}Suranaree University of Technology, Nakhon Ratchasima, Thailand
\item \Idef{org116}Technical University of Ko\v{s}ice, Ko\v{s}ice, Slovakia
\item \Idef{org117}Technische Universit\"{a}t M\"{u}nchen, Excellence Cluster 'Universe', Munich, Germany
\item \Idef{org118}The Henryk Niewodniczanski Institute of Nuclear Physics, Polish Academy of Sciences, Cracow, Poland
\item \Idef{org119}The University of Texas at Austin, Austin, Texas, United States
\item \Idef{org120}Universidad Aut\'{o}noma de Sinaloa, Culiac\'{a}n, Mexico
\item \Idef{org121}Universidade de S\~{a}o Paulo (USP), S\~{a}o Paulo, Brazil
\item \Idef{org122}Universidade Estadual de Campinas (UNICAMP), Campinas, Brazil
\item \Idef{org123}Universidade Federal do ABC, Santo Andre, Brazil
\item \Idef{org124}University of Cape Town, Cape Town, South Africa
\item \Idef{org125}University of Houston, Houston, Texas, United States
\item \Idef{org126}University of Jyv\"{a}skyl\"{a}, Jyv\"{a}skyl\"{a}, Finland
\item \Idef{org127}University of Liverpool, Liverpool, United Kingdom
\item \Idef{org128}University of Science and Technology of China, Hefei, China
\item \Idef{org129}University of South-Eastern Norway, Tonsberg, Norway
\item \Idef{org130}University of Tennessee, Knoxville, Tennessee, United States
\item \Idef{org131}University of the Witwatersrand, Johannesburg, South Africa
\item \Idef{org132}University of Tokyo, Tokyo, Japan
\item \Idef{org133}University of Tsukuba, Tsukuba, Japan
\item \Idef{org134}Universit\'{e} Clermont Auvergne, CNRS/IN2P3, LPC, Clermont-Ferrand, France
\item \Idef{org135}Universit\'{e} de Lyon, Universit\'{e} Lyon 1, CNRS/IN2P3, IPN-Lyon, Villeurbanne, Lyon, France
\item \Idef{org136}Universit\'{e} de Strasbourg, CNRS, IPHC UMR 7178, F-67000 Strasbourg, France, Strasbourg, France
\item \Idef{org137}Universit\'{e} Paris-Saclay Centre d'Etudes de Saclay (CEA), IRFU, D\'{e}partment de Physique Nucl\'{e}aire (DPhN), Saclay, France
\item \Idef{org138}Universit\`{a} degli Studi di Foggia, Foggia, Italy
\item \Idef{org139}Universit\`{a} degli Studi di Pavia, Pavia, Italy
\item \Idef{org140}Universit\`{a} di Brescia, Brescia, Italy
\item \Idef{org141}Variable Energy Cyclotron Centre, Homi Bhabha National Institute, Kolkata, India
\item \Idef{org142}Warsaw University of Technology, Warsaw, Poland
\item \Idef{org143}Wayne State University, Detroit, Michigan, United States
\item \Idef{org144}Westf\"{a}lische Wilhelms-Universit\"{a}t M\"{u}nster, Institut f\"{u}r Kernphysik, M\"{u}nster, Germany
\item \Idef{org145}Wigner Research Centre for Physics, Budapest, Hungary
\item \Idef{org146}Yale University, New Haven, Connecticut, United States
\item \Idef{org147}Yonsei University, Seoul, Republic of Korea
\end{Authlist}
\endgroup
  
\end{document}